\documentclass[aps,prl,tightenlines,superscriptaddress,twocolumn]{revtex4-1}
\usepackage{amsmath,amssymb,amsfonts,bm}
\usepackage{dcolumn}
\usepackage[colorlinks=true,linkcolor=blue,citecolor=blue]{hyperref}
\usepackage{tabularx}
\usepackage{booktabs}
\usepackage{colortbl}
\usepackage{graphicx}
\usepackage{diagbox,multirow}

\newcommand{\tabincell}[2]{\begin{tabular}{@{}#1@{}}#2\end{tabular}}

\begin{document}

\definecolor{mygray}{gray}{.9}
\definecolor{mypink}{rgb}{.99,.91,.95}
\definecolor{mygreen}{rgb}{0.819,0.933,0.933}

\title{Comments on ``Quantum theory cannot consistently describe the use of itself''}

\author{Liang Chen}
\affiliation{Mathematics and Physics Department, North China Electric Power University, Beijing, 102206, China}
\author{Ye-Qi Zhang}
\affiliation{Mathematics and Physics Department, North China Electric Power University, Beijing, 102206, China}

\begin{abstract}
Recently, a delicately designed Gedankenexperiment was proposed to check the self-consistence of quantum theory in the description of the agents who are using this theory. It was demonstrated that the quantum theory is inconsistent. Here a critical improvement is presented, which can lead to a consistent explanation of the Gedankenexperiment by using quantum theory.
\end{abstract}

\maketitle



We first give a brief introduction to the Gedankenexperiment proposed in Ref. \cite{Frauchiger2018NatComm}.
Suppose that F, $\overline{\rm F}$, W, and $\overline{\rm W}$ are four agents in the experiment, which may take many rounds.
Each round of the experiment is carried out as follows. There is a random number generator based on the measurement of a quantum system R, whose initial state is $|{\rm init\rangle_R}=\sqrt{1/3}|{\rm heads}\rangle_{\rm R}+\sqrt{2/3}|{\rm tails}\rangle_{\rm{R}}$. For each round of the experiment, Agent $\overline{\rm F}$ invokes this random number generator and sends a spin S to agent F. If agent $\overline{\rm F}$ got $|{\rm heads}\rangle_{\rm R}$, she sends  $|{\downarrow}\rangle_{\rm S}$ to agent F, else if she got $|{\rm tails}\rangle_{\rm R}$, she sends $|{\rightarrow}\rangle_{\rm{S}}=\sqrt{1/2}(|{\uparrow}\rangle_{\rm S}+|{\downarrow}\rangle_{\rm S})$. Then agent F measures spin S in the  orthogonal basis $\{|{\uparrow}\rangle_{\rm S},|{\downarrow}\rangle_{\rm S}\}$. Next, agent $\overline{\rm W}$ makes a measurement with respect to a basis containing the vector $|\overline{\rm ok}\rangle_{\rm \overline{L}}=\sqrt{1/2}\left(|{\rm \overline{h}}\rangle_{\rm \overline{L}}-|{\rm \overline{t}}\rangle_{\rm \overline{L}}\right)$ on the lab $\overline{\rm L}$. $\overline{\rm L}$ contains the agent $\overline{\rm F}$ and the quantum system R. Both $\overline{\rm L}$ and L (introduced in the following context) remain isolated during the experiment unless the protocol explicitly prescribes measurements applied to them. $|{\rm \overline{h}}\rangle_{\rm \overline{L}}$ and $|{\rm \overline{t}}\rangle_{\rm \overline{L}}$ are the orthogonal states corresponding to agent $\overline{\rm F}$'s measurement results $|{\rm heads}\rangle_{\rm R}$ and $|{\rm tails}\rangle_{\rm R}$, respectively. If agent $\overline{\rm W}$ got the result $\overline{\rm ok}$, $\overline{\rm w}=\overline{\rm ok}$, otherwise, $\overline{\rm w}=\overline{\rm fails}$. Then agent $\overline{\rm W}$ makes an announcement to agent W about his result. Then, agent W makes his own measurement with respect to a basis containing the vector $|{\rm ok}\rangle_{\rm L}=\sqrt{1/2}\left(\big|{-\frac{1}{2}}\rangle_{\rm L}-\big|{+\frac{1}{2}}\rangle_{\rm L}\right)$ on the lab L. L contains agent F and spin S at this stage. $\big|{-\frac{1}{2}}\rangle_{\rm L}$ and $\big|{+\frac{1}{2}}\rangle_{\rm L}$ are two orthogonal vectors corresponding to agent F's measurement results $|{\downarrow}\rangle_{\rm S}$ and $|{\uparrow}\rangle_{\rm S}$, respectively. If agent ${\rm W}$ got the result ${\rm ok}$, ${\rm w}={\rm ok}$, otherwise, ${\rm w}={\rm fails}$. Here ends one round of the experiment.

Analysing the results, one can find that the probability to get the result $(\overline{\rm w}, {\rm w})=(\overline{\rm ok}, {\rm ok})$ is 1/12 when one use the standard quantum theory to describe $\overline{\rm L}$ and L. However, if the agents make some logical reasoning based on their own measurement results at hand and three assumptions containing quantum theory, as shown in Ref. \cite{Frauchiger2018NatComm}, agent W will never get the result $\rm w=ok$ if he knows $\overline{\rm w}=\overline{\rm ok}$, i.e., the probability to get the result $(\overline{\rm w}, {\rm w})=(\overline{\rm ok}, {\rm ok})$ is zero. Hence, the analyzation arrives at a contradiction. The other two assumptions, except quantum theory, seem to be reasonable, so the two authors conclude that quantum theory cannot consistently describe the use of itself.

In this work, we show that there exists a critical improvement of the above logical reasoning. We assume that the three assumptions (Q), (S) and (C) given in Ref. \cite{Frauchiger2018NatComm} are correct, and use them in the following derivation. The key issue in the following derivation is that, for each round of the Gedankenexperiment, agent W uses quantum theory and actually had already acquired some information of lab L before he did the measurement on it. Hence, W should make conclusion at some time point based on the conclusions made by other agents at the same time point.

Let us first see how quantum theory works in the perspective of W. Suppose that at one round of the experiment, agent $\overline{\rm{W}}$ gets the result $\overline{\rm{w}}=\overline{\rm{ok}}$. Agent $\overline{\rm{W}}$ can make the statement that `I am certain that F knows that $\rm{z}=+\frac{1}{2}$ at time n:11', as shown in Table 3 of Ref. \cite{Frauchiger2018NatComm}. After receiving the announcement from agent $\overline{\rm{W}}$ that $\overline{\rm{w}}=\overline{\rm{ok}}$ at time n:21, agent W can also make the statement that `I am certain that F knows that $\rm{z}=+\frac{1}{2}$  at time n:11' by using the same derivation as agent $\overline{\rm{W}}$ has used, i.e., using assumptions (Q) and (C). Furthermore, agent W can make the statement that `I am certain that $\rm{z}=+\frac{1}{2}$  at time n:11' according to assumption (C). In addition, according to Eq. (6) in Ref. \cite{Frauchiger2018NatComm}, the measurement of $\overline{\rm{W}}$ at time n:20 does not change this certainty between time n:11 and n:30. So, at the time before the measurement with respect to a basis containing the vector $|\rm{ok}\rangle_{L}$, agent W has already got some information of state $| \rangle_{\rm L}$, he can confirm that $| \rangle_{\rm{L}}=|{+\frac{1}{2}}\rangle_{\rm{L}}$. Suppose that at time n:30, agent W did not make the measurement with respect to a basis containing the vector $|\rm{ok}\rangle_{L}$, he made the measurement in the orthogonal basis $\{|{-\frac{1}{2}}\rangle_{\rm{L}}, |{+\frac{1}{2}}\rangle_{\rm{L}}\}$, he would get the result $\rm{z}=+\frac{1}{2}$ definitely according to his information $\overline{\rm w}=\overline{\rm ok}$ and his knowledges (Q), (C), (S). So, when he makes the measurement with respect to a basis containing the vector $|\rm{ok}\rangle_{L}$, he has a probability $|\langle{{\rm{ok}}|{+\frac{1}{2}}}\rangle_{\rm L}|^2=1/2$ to get the result $\rm w=ok$. 

\begin{table*}[tb]
  \centering
  \caption{Time evaluation of quantum states. $|{\overline{\rm fails}}\rangle_{\overline{\rm L}}$ and $|{\rm fails}\rangle_{\rm L}$ are the quantum states orthogonal to $|{\overline{\rm ok}}\rangle_{\overline{\rm L}}$ and $|{\rm ok}\rangle_{\rm L}$, respectively, their explicit expressions are given by $|{\overline{\rm fails}}\rangle_{\overline{\rm L}}=\sqrt{1/2}\left(|{\overline{\rm h}}\rangle_{\overline{\rm L}}+|{\overline{\rm t}}\rangle_{\overline{\rm L}}\right)$, $|{\rm fails}\rangle_{\rm L}=\sqrt{1/2}\left(|{-\frac{1}{2}}\rangle_{\rm L}+|{+\frac{1}{2}}\rangle_{\rm L}\right)$. }\label{tab1}
  \renewcommand\arraystretch{1.2}
  \begin{tabular}{cc}
    \hline\hline
    \rowcolor{mygreen}
    operations given in Box 1 of Ref. \cite{Frauchiger2018NatComm} & quantum states after the operations \\
    \hline
    initialization of quantum system R & $\sqrt{\frac{1}{3}}|{\rm heads}\rangle_{\rm R}+\sqrt{\frac{2}{3}}|{\rm tails}\rangle_{\rm R}$ \\
    \rowcolor{mygray}
    $\overline{\rm F}$ measures R, sets the spin S & $\sqrt{\frac{1}{3}}|{\rm heads}\rangle_{\rm R}|{\downarrow}\rangle_{\rm S}+\sqrt{\frac{2}{3}}|{\rm tails}\rangle_{\rm R}|{\rightarrow}\rangle_{\rm S}$ \\
    $\overline{\rm F}$ sends spin S to F & $\sqrt{\frac{1}{3}}|{\overline{\rm h}}\rangle_{\overline{\rm L}}|{\downarrow}\rangle_{\rm S}+\sqrt{\frac{2}{3}}|{\overline{\rm t}}\rangle_{\overline{\rm L}}|{\rightarrow}\rangle_{\rm S}$ \\
    \rowcolor{mygray}
    F measures S & $\sqrt{\frac{1}{3}}|{\overline{\rm h}}\rangle_{\overline{\rm L}}|-{\frac{1}{2}}\rangle_{\rm L}+\sqrt{\frac{1}{3}}|{\overline{\rm t}}\rangle_{\overline{\rm L}}|{-\frac{1}{2}}\rangle_{\rm L}+\sqrt{\frac{1}{3}}|{\overline{\rm t}}\rangle_{\overline{\rm L}}|{+\frac{1}{2}}\rangle_{\rm L}$ \\
    $\overline{\rm W}$ measures lab $\overline{\rm L}$ &   \tabincell{c}{$|\overline{\rm ok}\rangle_{\overline{\rm L}}|{+\frac{1}{2}}\rangle_{\rm L}$ (probability$=1/6$) \\ $\sqrt{\frac{4}{5}}|{\overline{\rm fails}}\rangle_{\overline{\rm L}}|{-\frac{1}{2}}\rangle_{\rm L}+\sqrt{\frac{1}{5}}|{\overline{\rm fails}}\rangle_{\overline{\rm L}}|{+\frac{1}{2}}\rangle_{\rm L}$ (probability $=5/6$) }\\
    \rowcolor{mygray}
    W measures lab L  &  \tabincell{c}{ $|{\overline{\rm ok}}\rangle_{\overline{\rm L}}|{\rm ok}\rangle_{\rm L}$ (conditional probability $=1/2$, total probability $=1/12$) \\
                  $|{\overline{\rm ok}}\rangle_{\overline{\rm L}}|{\rm fails}\rangle_{\rm L}$ (conditional probability $=1/2$, total probability $=1/12$)   \\
                  $|{\overline{\rm fails}}\rangle_{\overline{\rm L}}|{\rm ok}\rangle_{\rm L}$ (conditional probability $=1/10$, total probability $=1/12$) \\
                  $|{\overline{\rm fails}}\rangle_{\overline{\rm L}}|{\rm fails}\rangle_{\rm L}$ (conditional probability $=9/10$, total probability $=3/4$)} \\
    \hline \hline
  \end{tabular}
\end{table*}

Now we reanalysis the probability related to the measurement result $(\overline{\rm w}, {\rm w})=(\overline{\rm ok}, {\rm ok})$ in the logical reasoning way. Obtaining this measurement result depends on four conditions step by step, $\rm r=tails$, ${\rm z}=+\frac{1}{2}$, $\overline{\rm w}=\overline{\rm ok}$, and $\rm w=ok$. Their conditional probabilities are $2/3$, $1/2$, $1/2$ and $1/2$, respectively. So we get $\frac{2}{3}\times\frac{1}{2}\times\frac{1}{2}\times\frac{1}{2}=\frac{1}{12}$, which is quantitatively consistent with the quantum theory result, Eq. (7) in Ref. \cite{Frauchiger2018NatComm}. Therefore, the contradiction proposed in Ref. \cite{Frauchiger2018NatComm} is inexistent.

In order to see how the logical reasoning in Ref. \cite{Frauchiger2018NatComm} breaks down, we focus on the inference from statements $\overline{\rm F}^{n:00}$ and $\overline{\rm F}^{n:01}$ to $\overline{\rm F}^{n:02}$, where the conclusion ${\rm w}={\rm fails}$ appears for the first time (see the discussion below Eq. (4) and Table (3) in Ref. \cite{Frauchiger2018NatComm}). This inference is conditional, and depends on the fact that the quantum state $\sqrt{1/2}\left(\big|{-\frac{1}{2}}\rangle_{\rm L}+\big|{+\frac{1}{2}}\rangle_{\rm L}\right)$ can not be changed before agent $\rm W$'s measurement on lab L at time n:30. However, the measurement of agent $\overline{\rm W}$ on lab $\overline{\rm L}$ at time n:20 breaks this condition. This is because labs L and $\overline{\rm L}$ are quantum entangled to each other by the spin S sent from agent $\overline{\rm F}$ to agent F, any measurement on lab $\overline{\rm L}$ will induce a collapse of both quantum states $| \rangle_{\overline{\rm L}}$ and $| \rangle_{\rm L}$ according to quantum theory. For completeness, in Table \ref{tab1}, we list the time evaluation of quantum states for each step of the experimental procedure given in Box 1 of Ref. \cite{Frauchiger2018NatComm}, these results are calculated by W using the standard quantum theory.

Based on the above discussion, our logical reasoning can be expressed explicitly as an Improved Assumption (C):

\textbf{Improved Assumption (C)}

Suppose that agent A has established that:

Statement $A^{\rm (i)}$:``I am certain that agent A$^{\prime}$, \emph{at the same time point as I make deduction}, upon reasoning within the same theory as the one I am using, is certain that $x=\xi$ at time $t$.''

Then agent A can conclude that:

Statement $A^{\rm (ii)}$:``I am certain that $x=\xi$ at time $t$.''

\begin{figure*}[tb]
  \centering
  \includegraphics[width=\textwidth]{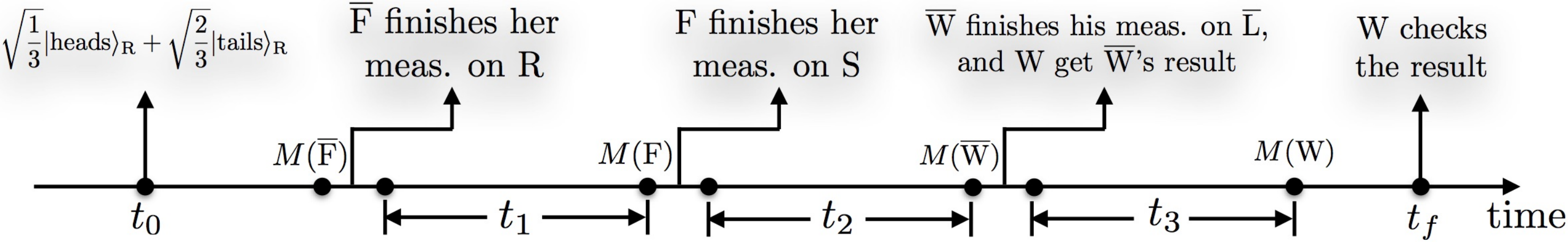}
  \caption{The flow chart of the whole process. $M(A)$ stands for the beginning of the measurement performed by agent A. For simplicity, we use $t_i$ represent both a time interval and an instant within this time interval, as illustrated in the picture. Once W receives $\overline{\rm W}$'s result $\overline{\rm w}=\overline{\rm ok}$, he makes a prediction on the probability of event $\textrm{w}=\textrm{ok}$ within the time interval $t_3$, based on quantum theory and logical reasoning.}
  \label{flow}
\end{figure*}

\begin{table}[tb]
	\centering
	\caption{Pathways of W's deduction through $\overline{\rm W}$, F, and $\overline{\rm F}$, there is 9 different combinations of time points.}\label{tab2}
	\begin{tabular}{|c|c|c|c|}
		\hline
		\diagbox{subject}{pathway}     & $\overline{\rm W}$ & F & $\overline{\rm F}$ \\\hline
		\multirow{4}*{$\mathrm{W(t_3)}$}& $t_1$ & $t_1$ & $t_1$\\\cline{2-4}
		& $t_2$ & $t_2$ & $t_2$\\\cline{2-4}
		& $t_3$ & $t_3$ & $t_3$\\
		\hline
	\end{tabular}
\end{table}

In order to see more clearly that this Improved Assumption (C) is consistence with Assumptions (Q) and (S), we would like to give some illustration below. The flow chart of the whole process is plotted in Picture \ref{flow}. We introduce here notations like $\overline{\rm F}(t_2){\rm F}(t_2)\overline{\rm W}(t_3){\rm W}(t_3)$ to represent a \emph{deduction} by W at $t_3$, through the conclusion by $\overline{\rm W}$ at $t_3$, through the conclusion by F at $t_2$, through the conclusion by $\overline{\rm F}$ at $t_2$. We will see it is crucial and reasonable to take account of these time points when someone deduces through another agent's conclusion. In the original paper, when $\overline{\rm W}$ obtain $\overline{\rm w}=\overline{\rm ok}$, the deduction $\overline{\rm F}(t_1){\rm F}(t_2)\overline{\rm W}(t_3){\rm W}(t_3)$ made by agent W at $t_3$ is ``At $t_3$, I am certain that, $\overline{\rm W}$ is certain at $t_3$ that, $\rm F$ is certain at $t_2$ that, $\overline{\rm F}$ is certain at $t_1$ that $\rm R=tails$ and $\rm w=ok$ will not happen''. We need to point out that the non-equal-time logical reasoning shown in the original paper \cite{Frauchiger2018NatComm} is inconsistent at the very beginning: agent $\overline{\rm W}$ gets the measurement result $\overline{\rm w}=\overline{\rm ok}$ at $t_3$ (in our notation, see Fig. \ref{flow}), he uses Eq. (6) in the original paper, the probability to get $z=-\frac{1}{2}$ is zero, to derive the result $|\rangle_{\rm L}=|{+\frac{1}{2}}\rangle_{\rm L}$ at $t_2$, furthermore, he deduces that $|\rangle_{\overline{\rm L}}=|{\overline{\rm t}}\rangle_{\overline{\rm L}}$ at $t_1$. However, if $|\rangle_{\overline{\rm L}}=|{\overline{\rm t}}\rangle_{\overline{\rm L}}$ at $t_1$, equation (6) in the original paper does not work. Straightforward calculation shows that the probability to get $z=-\frac{1}{2}$ is 1/2 in this case, not zero. The conclusion ($|\rangle_{\overline{\rm L}}=|{\overline{\rm t}}\rangle_{\overline{\rm L}}$) is conflicting with the condition (Eq. (6) in the original paper).

Furthermore, as shown in Table \ref{tab2}, there are 9 different deductions W could make at $t_3$. One can check that, different deductions can lead to different conclusions. There is no reason to abandon the other 7 non-equal-time deductions. The equal-time deduction listed in Table \ref{tab2} seems to be special and it can help us to avoid the contradictions induced by time back reasoning. We have shown, see Table \ref{tab1}, W's deduction $\overline{\rm F}(t_3){\rm F}(t_3)\overline{\rm W}(t_3){\rm W}(t_3)$ made at $t_3$ is ``at $t_3$, I am certain that, $\overline{\rm W}$ is certain at $t_3$ that, $\rm F$ is certain at $t_3$ that, $\overline{\rm F}$ is uncertain at $t_3$ that $\rm R=tail$ and $\rm w=ok$ will happen with some probability''. Now let us show the detailed calculation of this probability by logical reasoning based on the Improved Assumption (C) and check that the result coincides with that obtained by quantum theory.

Suppose at one round, saying round n, W receives the announcement from $\overline{\rm W}$ that $\overline{\rm w}=\overline{\rm ok}$ at time $t_3$, let us do the logical reasoning in W's perspective.

\underline{Step 1}: W is certain at $t_3$ that, $\overline{\rm W}$ is certain at $t_3$ that, (i) using the standard quantum theory, the probability to get this result ($\overline{\rm w}=\overline{\rm ok}$) is $1/6$, (ii) the state of lab $\overline{\rm L}$ at $t_3$ is
\begin{equation}
  | \rangle_{\overline{\rm L}}^{\overline{\rm W}\textrm{'s perspective}} = |\overline{\rm ok}\rangle_{\overline{\rm L}} = \frac{1}{\sqrt{2}}(|\overline{\rm h}\rangle_{\overline{\rm L}}-|\overline{\rm t}\rangle_{\overline{\rm L}}),
\end{equation}
and (iii) the state of lab L at $t_3$ is $| \rangle_{\rm L}^{\overline{\rm W}\textrm{'s perspective}}=|+\frac{1}{2}\rangle_{\rm L}$. In deducing point (iii), Equation (6) in Ref. \cite{Frauchiger2018NatComm} is used by agent $\overline{\rm W}$.

\underline{Step 2}: $| \rangle_{\rm L}=|+\frac{1}{2}\rangle_{\rm L}$ at $t_3$ means that agent F is certain at $t_3$ that the state of spin S is $|{\uparrow}\rangle_{\rm S}$. In F's perspective at $t_3$, labs $\overline{\rm L}$ and $\rm L$ are disentangled by her measurement on spin S and $\overline{\rm W}$'s measurement on $\overline{\rm L}$ cannot influence lab L. Hence, F at $t_3$ is certain that at $t_2$, $| \rangle_{\rm L}=|+\frac{1}{2}\rangle_{\rm L}$, furthermore, F at $t_3$ is certain that at $t_2$, $|\rangle_{\overline{\rm L}}=|\overline{\rm t}\rangle_{\overline{\rm L}}$. We (the authors of this correspondence) need to emphasize that, the improved assumption (C) has to be equal-time deduction. So agent F needs to know the state of $\overline{\rm L}$ at $t_3$, otherwise we cannot use the improved assumption (C). At any time, all of the agents know the experimental procedure presented in Box 1 of Ref. \cite{Frauchiger2018NatComm}, so agent F at $t_3$ knows that there is a measurement of $\overline{\rm W}$ on lab $\overline{\rm L}$. Using the standard quantum theory, F can get that $|{\overline{\rm t}}\rangle_{\overline{\rm L}}$ at $t_2$ would collapse to the mixed state of $|\overline{\rm ok}\rangle_{\overline{\rm L}}$ and $|\overline{\rm fails}\rangle_{\overline{\rm L}}$ with equal probability 1/2 at $t_3$ (F does not know $\overline{\rm W}$'s measurement result),
\begin{equation}
	| \rangle_{\overline{\rm L}}\langle |^{\textrm{F's perspective}}=\frac{1}{2}|{\overline{\rm ok}}\rangle_{\overline{\rm L}}\langle{\overline{\rm ok}}|+\frac{1}{2}|{\overline{\rm fails}}\rangle_{\overline{\rm L}}\langle{\overline{\rm fails}}|.
\end{equation}

\underline{Step 3.1}: At $t_3$, if $| \rangle_{\overline{\rm L}}^{\textrm{F's perspective}}=|{\overline{\rm ok}}\rangle_{\overline{\rm L}}$ (corresponding probability = $50\%$), agent $\overline{\rm F}$ can make the following logical reasoning. In her perspective at $t_3$: she knows that there is a measurement on lab $\overline{\rm L}$ at time between $t_2$ and $t_3$, however, just like the Wigner's Gedankenexperiment mentioned in Ref. \cite{Frauchiger2018NatComm}, in her perspective, her own state is not affected by this measurement and also the measurement of agent F on spin S, ``I am what I was''. So in her perspective at $t_3$, the state of lab $\overline{\rm L}$ which contains herself $|{\overline{\rm ok}}\rangle_{\overline{\rm L}}$ is not changed from $t_1$ to $t_3$. The event ``a spin S sent from lab $\overline{\rm L}$ to lab L'' is also superposition $\frac{1}{\sqrt{2}}\left(|\textrm{``sent a spin S = }{\downarrow}\textrm{''}\rangle-|\textrm{``sent a spin S = }{\rightarrow}\textrm{''}\rangle\right)$, hence the state of lab L after agent F's measurement is superposition, $| \rangle_{\rm L}^{\overline{\rm F}\textrm{'s perspective}}=\frac{1}{\sqrt{2}}\left[|\textrm{``L}=-{\frac{1}{2}}\textrm{''}\rangle^{\overline{\rm F}\textrm{'s perspective}}-|\textrm{``L}=\textrm{fails}_{\rm L}\textrm{''}\rangle^{\overline{\rm F}\textrm{'s perspective}}\right]$, where $|\textrm{``L}=\textrm{fails}_{\rm L}\textrm{''}\rangle$ is defined as $\frac{1}{\sqrt{2}}\left(|{-\frac{1}{2}}\rangle_{\rm L}+|{+\frac{1}{2}}\rangle_{\rm L}\right)$ and $|{\textrm{``L}=-\frac{1}{2}\textrm{''}}\rangle$ is defined as $|{-\frac{1}{2}}\rangle_{\rm L}$. We need to notify that the two states $|\textrm{``L}=\textrm{fails}_{\rm L}\textrm{''}\rangle$ and $|{\textrm{``L}=-\frac{1}{2}\textrm{''}}\rangle$ are not orthogonal, in order to get a normalized superposition, the normalization factor has to be changed from $\frac{1}{\sqrt{2}}$ to $\frac{1}{\sqrt{2-\sqrt{2}}}$. From agent $\overline{\rm F}$'s perspective, this state is not changed by agent $\overline{\rm W}$'s measurement, so from the superposition $\overline{\rm F}$'s point of view, the quantum state of lab L at $t_3$ is,
\begin{eqnarray}
&| \rangle_{\rm L}^{\overline{\rm F}\textrm{'s perspective}}= \frac{1}{\sqrt{2-\sqrt{2}}} \notag \\
&\left[|\textrm{``L}=-{\frac{1}{2}}\textrm{''}\rangle^{\overline{\rm F}\textrm{'s perspective}}-|\textrm{``L= fails}_{\rm L}\textrm{''}\rangle^{\overline{\rm F}\textrm{'s perspective}}\right].
\end{eqnarray}
Under this condition, at $t_3$, $\overline{\rm F}$'s prediction of probability of $\textrm{w}=\textrm{ok}$ is,
\begin{eqnarray}
&|\textrm{``the probability of `} \textrm{w}=\textrm{ok' ''}\rangle^{\overline{\rm F}\textrm{'s perspective}}= \frac{1}{\sqrt{2-\sqrt{2}}} \notag \\
&\left[|\textrm{``the probability of `} \textrm{w}=\textrm{ok'}=\frac{1}{2}\textrm{''}\rangle^{\overline{\rm F}\textrm{'s perspective}}-\right. \notag \\
&\left.|\textrm{``the probability of `} \textrm{w}=\textrm{ok'}=0\textrm{''}\rangle^{\overline{\rm F}\textrm{'s perspective}} \right]. \label{eq4}
\end{eqnarray}
From agent F's perspective, agent $\overline{\rm F}$ in the superposition state of lab $\overline{\rm L}$ does not have the ability to give a definite conclusion about what the probability of $\textrm{w}=\textrm{ok}$ is: i.e., if she has such an ability to make a statement, e.g., ``I (agent $\overline{\rm F}$) am certain that the probability of $\textrm{w}=\textrm{ok}$ is 1/2'', which demonstrates that $\overline{\rm L}$ is not superposition, $|\rangle_{\overline{\rm L}}=|{\overline{\rm t}}\rangle_{\overline{\rm L}}$ in this case, and conflicts with agent F's judgement of lab $\overline{\rm L}$ at $t_3$. However, agent F knows that $|\rangle_{\overline{\rm L}}=|\overline{\rm ok}\rangle_{\overline{\rm L}}$, so if she opens the lab $\overline{\rm L}$ at $t_3$, there is a probability $\left(\frac{1}{\sqrt{2-\sqrt{2}}}\right)^2=\frac{1}{2-\sqrt{2}}$ to receive the message ``the probability of $\rm{w}=\rm{ok}$ is 1/2'' from agent $\overline{\rm F}$, and a probability $\frac{1}{2-\sqrt{2}}$ to receive the message ``the probability of $\rm{w}=\rm{ok}$ is 0''. So, without confusion, we can effectively say that ``(From agent F's perspective at $t_3$) agent $\overline{\rm F}$ is certain at $t_3$ that the probability to get $\rm{w}=\rm{ok}$ is $\frac{1}{2-\sqrt{2}}\times\frac{1}{2}+\frac{1}{2-\sqrt{2}}\times0$'' in this situation.


\underline{Step 3.2}: If $| \rangle_{\overline{\rm L}}^{\textrm{F's perspective}}=|{\overline{\rm fails}}\rangle_{\overline{\rm L}}$ (corresponding probability = $50\%$), agent $\overline{\rm F}$ can use a similar logical reasoning and get that the probability to get ${\rm w}={\rm ok}$ is $\frac{1}{2+\sqrt{2}}\times\frac{1}{2}$.

All of the considerations presented above are carried out at time $t_3$, and we (the authors of this correspondence) need to emphasize that all of the above considerations are essentially carried out in agent W's mind. Now agent W can make the deduction using the Improved Assumption (C):

Statement $A^{\rm (i)}$: ``I am certain at $t_3$ that, $\overline{\rm W}$ is certain at $t_3$ that, F is certain at $t_3$ that, $50\%$ probability that $\overline{\rm F}$ is certain at $t_3$ that the probability to get ${\rm w}={\rm ok}$ is $\frac{1}{2-\sqrt{2}}\times\frac{1}{2}$ and $50\%$ probability that $\overline{\rm F}$ is certain at $t_3$ that the probability to get ${\rm w}={\rm ok}$ is $\frac{1}{2+\sqrt{2}}\times\frac{1}{2}$.''

So agent W can make the following deduction:

Statement $A^{\rm (ii)}$: ``I am certain at $t_3$ that the probability to get ${\rm w}={\rm ok}$ is $50\%\times\frac{1}{2-\sqrt{2}}\times\frac{1}{2}+50\%\times\frac{1}{2}\times\frac{1}{2+\sqrt{2}}\times\frac{1}{2}=\frac{1}{2}$.''

The probability to get $\overline{\rm w}=\overline{\rm ok}$ is 1/6, so agent W can conclude that the total probability to get $({\rm w}, \overline{\rm{w}})=({\rm ok}, \overline{\rm{ok}})$ is 1/12.

So we conclude that the Improved Assumption (C) can handle the contradiction discussed in Ref. \cite{Frauchiger2018NatComm}, i.e., the inconsistence of direct calculation using quantum theory and indirect calculation based on logical reasoning.

\underline{Further Remarks}: Among the above derivations, the logical reasoning shown in Step 3.1 may make one being confused: what are the thoughts and logical reasonings of an agent who is in a superposition state? We need to emphasize that: (i) when we use a superposition state $|{\overline{\rm ok}}\rangle_{\rm L}$ to describe an isolated lab $\overline{\rm L}$ with an agent $\overline{\rm F}$ contained (like that considered in Ref. \cite{Frauchiger2018NatComm} and references therein), we have to accept that every thing in the isolated lab $\overline{\rm L}$ should be superposition, e.g., agent $\overline{\rm F}$'s notes written on a paper in lab $\overline{\rm L}$ to remind herself in future or other observer about her current measurement results, her memory about the history,  her judgement of her current situation, her prediction of the future. (ii) such a strange state is unobservable (at least for the current experiment). Suppose at some time $|\rangle_{\overline{\rm L}}=|{\overline{\rm ok}}\rangle_{\overline{\rm L}}$, agent $\overline{\rm W}$ wants to see what this state is, and more incisively, he wants $\overline{\rm F}$ herself to tell him that she is in a superposition state, then, he will find that, when he open the lab, the state $|{\overline{\rm ok}}\rangle_{\overline{\rm L}}$ collapses to $|{\overline{\rm h}}\rangle_{\overline{\rm L}}$ or $|{\overline{\rm t}}\rangle_{\overline{\rm L}}$ immediately and everything is normal and non-superposition. This is because the pre-arranged experiment procedure limited the projection operator (open the lab) to be $\left\{|{\overline{\rm h}}\rangle_{\overline{\rm L}}\langle{\overline{\rm h}}|,|{\overline{\rm t}}\rangle_{\overline{\rm L}}\langle{\overline{\rm t}}|\right\}$. (iii) we have to admit that the two quantum states, $|\textrm{``the probability of `} \textrm{w}=\textrm{ok'}=\frac{1}{2}\textrm{''}\rangle$ and $|\textrm{``the probability of `} \textrm{w}=\textrm{ok'}=0\textrm{''}\rangle$ are not orthogonal. Because the effective dimension of Hilbert space is 4 in this problem. If different values in $[0,1]$ for the probability of $\textrm{w}=\textrm{ok}$ refer to different (orthogonal) quantum states, then the dimension of the Hilbert space will be infinite. This discuss guarantees that the prefactor $\frac{1}{\sqrt{2-\sqrt{2}}}\ne\frac{1}{\sqrt{2}}$ is reasonable.

\end{document}